# Asynchronous Transmission of Wireless Multicast System with Genetic Joint Antennas Selection

Jihua Lu, Xiangming Li, Dan Liu, Xiangyuan Bu
School of Information and Electronics
Beijing Institute of Technology
Beijing, China
e-mail: lujihua@bit.edu.cn

*Abstract*—Optimal antenna selection algorithm of multicast transmission can significantly reduce the number of antennas and can acquire lower complexity and high performance which is close to that of exhaustive search. An asynchronous multicast transmission mechanism based on genetic antenna selection is proposed. The computational complexity of genetic antenna selection algorithm remains moderate while the total number of antennas increases comparing with optimum searching algorithm. Symbol error rate (SER) and capacity of our mechanism are analyzed and simulated, and the simulation results demonstrate that our proposed mechanism can achieve better SER and sub-maximum channel capacity in wireless multicast systems.

*Keywords-Multicast system; generic antenna selection; asynchronous transmission; symbol error rate; system capacity*

I. INTRODUCTION

Enhanced multimedia broadcast multicast service (EMBMS) of the long term evolution (LTE) standard can markedly boost spectral efficiency and reduce the costs per bit when the same content must be delivered wirelessly to multiple mobile users. The goal of multicast transmission is to minimize the total transmission power at the Base Station (BS) while guaranteeing a certain minimum signal-to-noise ratio (SNR) per user[1]. On one hand, the high resolution radio frequency chain costs high; the antenna elements, on the other hand, are becoming smaller and cheaper; thus, antenna selection strategies are becoming increasingly desirable. Mobile multicast broadband spectrum services, e.g. high-speed traffic detection and online video conferences, attract lots of attention, due to its flexibility and efficient data transmission from one source to several independent receivers. Transmit-diversity has been widely accepted to combat deep channel fading and has been mainly used in point-to-point wireless communications. However, the employment of transmit-diversity in mobile multicast (point-to-multipoint) systems hasn't been thoroughly studied[2]. The closed-loop transmit-diversity scheme requires Channel State Information (CSI) and can achieve better performance in multicast systems[1][2]. Classical antenna selection methods are based on the maximum likelihood (ML) criterion or minimizing SER. However, such approaches cannot guarantee the optimum effectiveness of the antenna selection as the optimum antenna subset does not necessarily yield ideal performances[3].

A decoupled antenna selection scheme was proposed by A. Gorokhov and M. Collados, which has less complexity and reductive computational burden of obvious performance degradation through combining with trans/receive antenna selection[4][5]. Moreover, some efficient antenna selection schemes with relatively large performance degradation have been proposed in [6] and [7], and a hybrid genetic antenna selection algorithm was proposed in [8]. Since the Priority-Based Genetic Algorithm (PGA) has preferable SER performance, however, all the priorities in which are integers, thus two antennas may own the same priority after crossover, and it will need a sort-refill operation subsequently[6]. Jenn-Kaie Lain extended the PGA to Real-Valued Genetic Approach (RVGA) through setting the priority be real value variables but not integers only, which improves the algorithm performance and remains a low computational complexity by reducing the exchange-empty and sort-refill operations in PGA[3]. The performances of RVGA method are also inferior to the exhaustive antenna selection algorithm. We proposed a full weighted cross-genetic antenna selection algorithm in mobile wireless communication systems based on multicast transmission scenarios, which could achieve almost maximum channel capacity with affordable complexity and better SER performance.

The rest of our paper is organized as follows. The system model of multicast system with antenna selection scheme is described in Section II. The genetic algorithm in multicast system was proposed in Section III. The simulation results of system SER and capacity are stated in Section IV and followed by the conclusions in Section V.

II. MULTICAST SYSTEM WITH ANTENNA SELECTION SCHEME

*A. Multicast system model with antenna selection*

Consider a downlink multicast system with antenna selection, which includes one base station and R mobile receivers. As shown in Fig.1, the sender transmits the same common data source to R mobile receivers.

We analyze a mobile wireless link comprising M transmit and N receive antennas in a Rayleigh fading environment, and the channel gain matrix can be expressed as $H = [h_1^T h_2^T \cdots h_R^T]^T$, where $(\cdot)^T$ is the matrix transpose and $h_r = (h_{n,m}^{(r)})_{N \times M}$ ($r = 1, 2, \cdots, R; n = 1, 2, \ldots, N; m = 1, 2, \ldots, M$) stands for the channel gain between the $m$-th transmit antenna and the $n$-th receive antenna of the $r$-th mobile receiver. We assume block channel between the individual transmit-receive antenna pairs stays constant during one signal block, but varies every signal block.

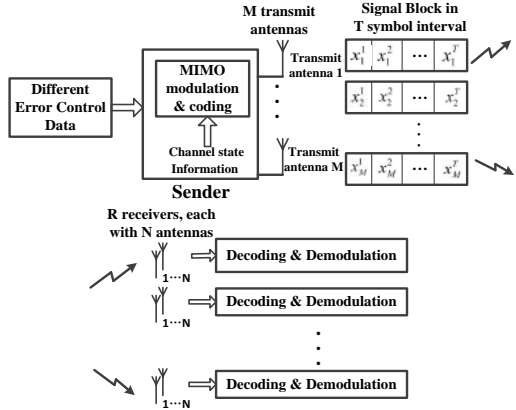

Figure 1. Multicast transmission system based on antenna selection

### B. Discrete Channel Model of frequency-selective Multicast Systems

For a given delay, elements in channel matrix subject to Rayleigh distribution and have the same average power, which is denoted as $P = E(|h_{n,m}^{(r)}|^2)$ and suited for each $h_{n,m}^{(r)}$. The correlation coefficients between any two delays equal to zero (which means any two $h_{n,m}^{(r)}$ parameters are irrelevant), and the cross correlation of $h_{n,m}^{(r)}$ with two different delays can be expressed as:

$$\langle h_{n,m}^{(r)\tau_1}, h_{n,m}^{(r)\tau_2} \rangle = 0 \quad \tau_1 \neq \tau_2 \quad (1)$$

Where, the symbol $\langle , \rangle$ represents computing the correlation coefficient of two variables. The Average Power Delay (APD) can be expressed as

$$P(\tau) = \sum_{l=1}^{L} P_l \delta(\tau - \tau_1) \quad (2)$$

By suitably setting the time delay and average power parameters $\{\tau_l, P_l\}$, we can achieve the APD with some specific time delay expansion.

Assume each signal block is denoted by matrix $X = (x_m^t)_{M \times T}$, where $x_m^t$ is baseband symbol transmitted from the $m$-th transmit antenna during the $t$-th symbol interval. And assume the transmit antenna-weight matrix is $W_r \triangleq diag(w_1^{(r)}, w_2^{(r)}, \ldots, w_M^{(r)})$, with $W_r^* W_r = 1$, where * denotes conjugate operation. Then, the transmitted signal can be written as $W_r X$.

Thus, the discrete MIMO channel model with frequency selective fading can be expressed as

$$y_r = h_r W_r X + n_r = (h_{n,m}^{(r)})_{N \times M} W_r X + n_r \quad (3)$$

Where $n_r$ is zero mean, complex Additive Gaussian White Noise (AWGN) matrix of the $r$-th mobile receiver with size of $N \times T$. Assume $Y = [y_1^T, y_2^T, \cdots, y_R^T]^T$, $W = [W_1, W_2, \cdots, W_R]^T$ and $Z = [n_1^T, n_2^T, \cdots n_R^T]^T$, then (3) can be simplified as $Y = HWX + Z$.

### III. GENETIC ANTENNA SELECTION AND ITS PERFORMANCES IN MULTICAST SYSTEM

#### A. Genetic Antenna Selection in Multicast System

In multimedia broadcast multicast service (MBMS) of LTE networks, efficient power allocation techniques e.g. antenna selection should be implemented so as to ensure the system capacity and the massive multimedia transmission requirements between mobile users[9]. However, even multicast beamforming with antenna selection is a NP-hard problem[1][10]. Assume there are $L_S$ and $L_U^{(r)}$ RF chains at the transmitter and receiver sides, respectively, and $L_U^{(r)}$ corresponds the $r$-th receiver. Antenna selection algorithm selects $L_S \times L_U^{(r)}$ out of $M \times N^{(r)}$ antennas and finds the corresponding weight vector $\overline{W_r}$ (subset of $W_r$) to maximize the system capacity through minimizing the transmit-power, subject to the receive-SNR constraints per receiver[8]. The flow chart of the genetic antenna selection algorithm is shown in Fig. 2. We apply antenna selection to determine an appropriate antenna subset from all antennas with the maximum capacity can be expressed as

$$\overline{C} = \log_2 \det(I_{L_S} + \frac{E_S}{L_S N_0} \overline{H}^{(r)H} \overline{H}^{(r)}) \quad (4)$$

Where $E_s$ is the power of the transmitted signal per symbol, $N_0$ is the unilateral noise power spectral density, $\overline{H}^{(r)}$ is a $L_S \times L_U^{(r)}$ sub-matrix of $h_r$. From (3) and (4), it is easy to make conclusion that the optimal antenna selection could be obtained through exhaustive searching over all the possible antennas' matched pairs. However, the exhaustive search takes an unaffordable computational load especially when the number of antennas is quite large. Thus, many researchers work on developing suboptimal antenna selection algorithm with less computational complexity. We present the flow chart of genetic antenna selection in multicast system in Fig. 2, which contains the Initialize step, conventional step, priority-based step, random mutation step and et. al.

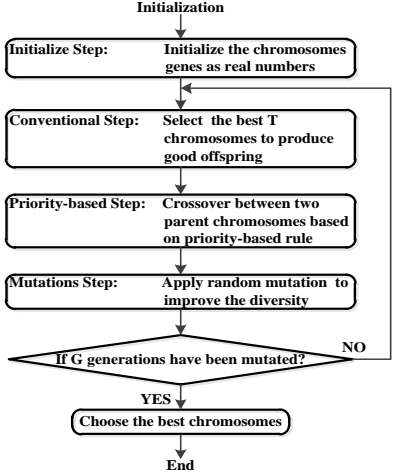

Figure 2. Flow chart of Genetic Algorithm in a multicast system

First, we define an initial population of Q (half to double of the maximum receiving antennas' number) chromosomes in **the initialize step** and each chromosome contains $(N^{(r)}+M)$ genes. Suppose $p$ is the priority of each gene which is a zero-mean Gaussian random variable with the variance of $\sigma^2$. Second, the chromosomes with highest scores (value of capacity function which is shown in Eq.(4)) be chosen in **the conventional step** to select the ones to put into the mating pool where they will be mated randomly. Third, all the parent chromosomes' genes have been crossed over at a random rate rather than exchange parent chromosomes' genes based on the crossover mask in **the priority-based step**. Four, we use a mutation mask, and the elements in which are 1s with the probability of $p_m$ and 0s with the probability of $1-p_m$ in **the mutation step**. And owing to the crossover operation in the conventional step compromise with two priority magnitudes to get new samples, therefore we enlarge the mutation amplitude to gain more diversity. Repeat **the conventional step**, **priority-based step** and **mutation step** to produce **G** generations of chromosomes. Finally, we choose the chromosomes which own the highest scores in the **G**-th generation. In **the end**, $L_S$ genes with the highest values out of the transmitting $M$ genes and the $L_U^{(r)}$ genes with the highest values out of the $N$ genes which belong to corresponding user will be chosen and one complete antenna selection process is finished.

### B. Capacity of Genetic Antenna Selection in a Wireless Multicast System

The score function to conduct in the conventional selection step of our presented algorithm is the elitism operation. We assign real value variables to be the priority of each antenna in our presented method in a wireless multicast system. So the system capacity formula Eq.(4) can be reformulated as

$$\overline{C} = \max_{p \in \mathbb{R}^{M+N^{(r)}}} (\log_2 \det(I_{L_S} + \frac{E_S}{L_S N_0} \overline{H}^{(r)}(p)^H \overline{H}^{(r)}(p))) \quad (5)$$

$\overline{H}^{(r)}(p)$ is a sub-block matrix with $L_U^{(r)}$ rows and $L_S$ columns which are selected from the channel matrix $h_r$ in Eq.(3) with the same connotation defined as $\overline{H}^{(r)}$ in Eq. (4). The priority magnitude vector (PMV) of $\overline{H}^{(r)}(p)$ is $P = [p_1^{(r)}, p_2^{(r)}, ..., p_N^{(r)}, p_{N+1}, ..., p_{N+M}]$, which means the priority vector corresponding to $N^{(r)}$ receive antennas and $M$ transmit antennas. The receive antennas with $L_U^{(r)}$ largest values out of the first $N^{(r)}$ elements in **P** are selected as the rows in $\overline{H}^{(r)}(p)$ while the transmit antennas with $L_S$ largest values out of the last $M$ elements in P are selected as the columns in $\overline{H}^{(r)}(p)$.

### C. SER of Genetic Antenna Selection in a Wireless Multicast System

Symbol Error Rate (SER) is used as performance metrics in the downlink multicast system described in Fig. 1. Define the average SER for the $r$-th receiver is

$$\overline{p}_r = \int_0^\infty f_{R_r}(\gamma) p(\gamma) d\gamma \quad (6)$$

Where $f_{R_r}(\gamma)$ is the SER for the K-ary rectangle Quadrature Amplitude Modulation (QAM) in AWGN channels with the SNR $\gamma$ and $p(\gamma)$ is the probability density function (pdf) of the combined SNR $\gamma_r$ of the $r$-th receiver.

添加公式:

$$f_{R_r}(\gamma) = 4(1 - \frac{1}{\sqrt{M}}) Q(\sqrt{\frac{3}{M-1} \cdot \frac{E_b}{N_o} \cdot \gamma^2}) \quad (7)$$

where $Q(p) \triangleq \frac{1}{\sqrt{2\pi}} \int_p^\infty \exp(-\frac{u^2}{2}) du$ is known as the Q-function.

The pdf of the combined SNR for hybrid selection-MRC in Rayleigh fading channels is expressed as following:

$$p(\gamma) = \frac{\gamma^{L-1} e^{-\gamma/\overline{\gamma}}}{(L-1)! \overline{\gamma}^L} \quad (8)$$

where L is the number of available paths.

Then applying (7) and (8) to (6), we can get the average SER for the r-th receiver is:

$$\overline{p}_r = \int_0^\infty (4(1 - \frac{1}{\sqrt{M}}) Q(\sqrt{\frac{3}{M-1} \cdot \frac{E_b}{N_o} \cdot \gamma^2})) (\frac{\gamma^{L-1} e^{-\gamma/\overline{\gamma}}}{(L-1)! \overline{\gamma}^L}) d\gamma \quad (9)$$

我把这段放下面了。Assume the average transmit power per symbol, denoted by $\overline{S}$ is equal to normalized constant 1. Thus, the average received SNR per symbol per receive antenna $\overline{R}_{ac}$ can be expressed as $\overline{R}_{ac} = \overline{S}/\sigma^2$. In each mobile receiver, we apply selection combing algorithm to combine signals received from different antennas.

这里需要完善公式(6)，代入得公式(7).另外没有下文展开，感觉太单薄了。加个一句两句的会丰满一些。

## IV. Performances of Genetic Antenna selection in Wireless Multicast System

In this section, we evaluate the SER and capacity performances of an asynchronous transmission multicast system using priority-based generic algorithm via simulations. Consider the downlink multicast system described in Fig. 1, we choose 16QAM for data transmission asynchronously through $L_T$ out of $M$ antennas in independent identically distributed (i.i.d.) Rayleigh fading channels. Each channel matrix element $h_{n,m}^{(r)}$ is modeled as a circularly symmetric complex Gaussian random variable with variance equals to constant 1, which is also known as Rayleigh fading.

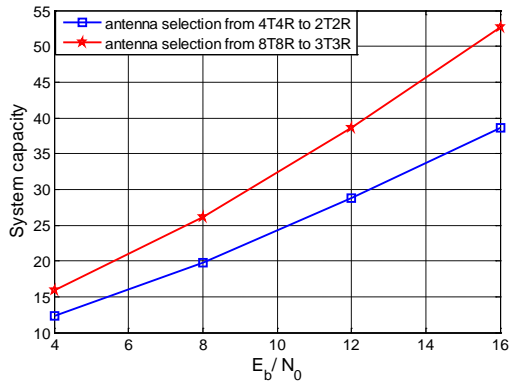

Figure 3. Capacity performance of different antenna selection scheme in multicast system

Fig. 3 shows the capacity performance of multicast system with antenna selection algorithm. We define "xTyR" as "x" transmit antennas and "y" receive antennas, so "4T4R to 2T2R" in the first curve of square line demonstrates selecting two transmit and two receive antennas from four transmit antennas (M=4) and four receive antennas in the presented four receivers' multicast system. In this antenna selection scheme, we choose the population size is P=20, chromosomes in mating pool are T=8, generation number G=12, mutation probability Pm=0.09, and crossover probability is Pc=0.75. "8T8R to 3T3R" in the second curve of star line demonstrates selecting three transmit and three receive antennas from eight transmit antennas (M=8) and eight receive antennas in the presented four receivers' multicast system. In the second antenna selection scheme, the population size is P=40, chromosomes in mating pool are T=16, generation number G=24, mutation probability Pm=0.09, and crossover probability is Pc=0.75. As shown in Fig. 3, the first scheme selects three transmit/receive antennas from 8T8R (eight transmit and eight receive antennas) for each receiver has better capacity performance than the second scheme. In addition, due to the asynchronous transmission, system capacity in a multicast system is significantly improved.

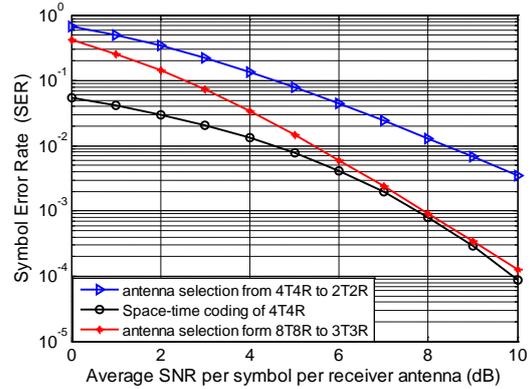

Figure 4. SER performance in a multicast wireless system between different antenna selection schemes

Fig. 4 is the comparison between the average SNR per receive antenna of different antenna selection schemes and space-time coding in a multicast asynchronous transmission. Each receiver exploits generalized selection combining to gather the corresponding signals received from different antennas. As discussed in Section III, we use Eq. (6) to calculate the symbol error rate, where the pdf of generalized selection combining-MRC in Rayleigh fading environments have been derived in [12]. From Fig. 4, we observe that antenna selection scheme of 8T8R to 3T3R can achieve the approximate performance of simple space-time coding of 4T4R, and this demonstrates that antenna selection scheme can significantly enhance the system reliability besides its high capability.

## V. Conclusions

A downlink multicast scenario with genetic antenna selection scheme is presented. The sender is equipped with multiple antennas broadcasting successive data packets formed in groups to several users with multiple antennas over a common bandwidth. The goal of antenna selection is to select appropriate weight vectors so as to maximize the minimum signal-to-interference-plus-noise ratio (SINR) among all users under a power constraint. The proposed antenna selection algorithm can improve the system capacity significantly with lower complexity and is close to that obtained through optimal search.

### Acknowledgment

This work was supported in part by NSF of China with grants 61002014, 60972017 and 60972018, 61072048, the Important National 863 Specific Projects with grant 2012AA121604, the Excellent Young Teachers Program of MOE, PRC with grant 2009110120028, and the Research Fund for the Doctoral Program of Higher Education with grants 20091101110019.